\documentclass[pra, twocolumn,notitlepage,superscriptaddress,nofootinbib,nolongbibliography]{revtex4-1}
\usepackage[
    colorlinks=true,
    linkcolor=black,
    citecolor=black,
    urlcolor=black
]{hyperref}
\usepackage{braket}
\usepackage[dvipsnames]{xcolor} %
\usepackage{amsmath, amssymb, graphicx, bm}
\usepackage{braket}
\usepackage{comment}
\usepackage{tikz,pifont}
\usepackage{color}
\usepackage{dsfont}
\usepackage{orcidlink}
\usepackage[normalem]{ulem}

\newcommand{\id}{\mathbb{I}}

\newcommand{\zm}{\mathbf{0}}
\begin{document}
	\title{Parameter-Shift Rules for Gradients in Boson Sampling Experiments}%
    \author{Marius Trudeau \textsuperscript{\orcidlink{0009-0000-0273-9894}}}
    \email{marius.trudeau@etud.polymtl.ca}
    	\affiliation{%
    	D\'epartement de g\'enie physique, \'Ecole polytechnique de Montr\'eal, Montr\'eal, QC, H3T 1J4, Canada 
    }
    \author{Pierre-Emmanuel Emeriau \textsuperscript{\orcidlink{0000-0001-5155-1783}}}
    \email{pe.emeriau@quandela.com}
    \affiliation{\href{https://ror.org/01yjzw780}{Quandela SAS}, 7 Rue L\'eonard de Vinci, 91300 Massy, France}
    	\author{Nicol\'as Quesada\textsuperscript{\orcidlink{0000-0002-0175-1688}}}
	\email{nicolas.quesada@polymtl.ca}
	\affiliation{%
		D\'epartement de g\'enie physique, \'Ecole polytechnique de Montr\'eal, Montr\'eal, QC, H3T 1J4, Canada
	}

	\date{\today}

	\begin{abstract}
        Using a robust photon loss model for photonic quantum experiments, we derive $n-$th order parameter-shift rules for computing gradients of Fock boson sampling transition probabilities where $n$ photons are sent into a lossy interferometer.
        We also show that in general it is not possible to generalize this gradient recipe to the case of Gaussian boson sampling. Only in the specific case where the transmission matrix of the interferometer can be factorized as a diagonal loss matrix premultiplied by a pure unitary it is possible to obtain parameter-shift rules with a finite order given by twice the total number of photons detected.
        We demonstrate the efficacy of the proposed method by comparing its performance against finite differences on real hardware.%
    \end{abstract}

    		\maketitle
    \section{Introduction} 
    
    Integrated photonic systems offer significant potential for near-term quantum computing. Photonic chips, in particular, are scalable due to their use of well-optimized micro-fabricated components and can exhibit lower decoherence than matter-based qubits~\cite{rudolph2017optimistic, su2023scalability, pelucchi2022potential}. Another significant advantage is that they can operate at room temperature, reducing the need for the complex cryogenics required by many other quantum computing platforms~\cite{su2023scalability, rudolph2017optimistic, ashtiani2026integrated}. Recent experimental progress further highlights the promise of this technology. Ongoing industrial efforts continue to demonstrate ever larger programmable, and robust photonic computers capable of implementing increasingly complex protocols~\cite{psiquantum2025manufacturable, aghaee2025scaling, maring2024versatile}. Integrated photonics is rapidly maturing and progressing toward scalable quantum processors. Such devices represent an important step toward achieving quantum advantage and realizing a fault-tolerant quantum computer~\cite{rudolph2017optimistic, pelucchi2022potential, arrazola2021quantum, wang2020integrated, zhu2026recent}. 
    
    Despite their promise and demonstrated success in multiple applications, such as quantum sensing~\cite{pirandola2018advances} and quantum communication~\cite{lo2014secure}, loss remains a critical challenge in photonic systems~\cite{rudolph2017optimistic, pelucchi2022potential, arrazola2021quantum, wang2020integrated, zhu2026recent}. Photon loss can originate from several sources, including imperfect coupling between optical components, propagation losses in waveguides, imperfect photon sources and detectors, and fabrication imperfections in integrated photonic circuits~\cite{clements2016optimal, poveda2025etching, bombardelli2025foundations, dong2023programmable, wang2008wideband, reed2010silicon}. Loss is particularly detrimental in quantum photonic systems since it directly destroys information by reducing the number of photons participating in a computation, hindering interference in the process~\cite{yao2024riemannian, bulmer2022threshold, bulmer2026simulating}. To circumvent this, post-selection has been used to only consider output events where they obtain perfect transmission. Although this guarantees that no loss has occurred in the retained events, this method discards a majority of the output data due to photon losses, leading to extremely inefficient data usage. Since the probability of perfect transmission decreases as the number of modes and optical components increases, this approach quickly becomes infeasible for large-scale photonic systems~\cite{lund2017quantum, wang2018toward}. Thus, as the size of photonic quantum processors increases, accurately modeling losses becomes more important for benchmarking, optimizing algorithms, and developing error mitigation and correction schemes. Accurate mathematical modeling of photon losses is essential for the efficient implementation of algorithms on real photonic devices. 
    
    A common task required for a multitude of quantum algorithms is optimization. %
    Variational Quantum Algorithms (VQAs) have emerged as one of the most promising avenues for near-term quantum computing, as they outsource the parameter optimization process to classical computers while still using quantum system observables~\cite{cerezo2021variational, qi2024variational}. VQAs encompass a broad family of algorithms, including Variational Quantum Eigensolvers (VQEs)~\cite{peruzzo2014variational, tilly2022variational}, Quantum Approximate Optimization Algorithms (QAOAs)~\cite{farhi2014quantum, blekos2024review}, variational quantum classifiers (VQCs)~\cite{schuld2019circuit, havlicek2019supervised}, quantum neural networks (QNNs)~\cite{killoran2019continuous, schuld2014quest, schuld2021machine, zhou2023quantum, jeswal2019recent, altaisky2001quantum}, as well as variational quantum simulation (VQS)~\cite{li2017efficient, yuan2019theory}. In all of these cases, efficiently calculating gradients with respect to circuit parameters is highly relevant. 

    Several approaches exist to estimate these gradients. The most straightforward technique is the Finite Difference (FD) method, which estimates derivatives from small shifts around the parameter values~\cite{teo2023optimized, liang2025qugstep}. However, this method is extremely sensitive to noise and the choice of the FD step size, making this method particularly unstable on noisy systems. 

    In this manuscript, a method is proposed to exactly calculate the derivative of transition probabilities using only system observables in the presence of photon loss. In particular, the proposed method belongs to a class of algorithms commonly referred to as parameter-shift rules (PSRs)~\cite{pappalardo2024photonic, facelli2024exact, wierichs2022general}. These methods are based on theoretically exact expressions of derivatives and mitigate the sensitivity of the FD technique to phase-shift imprecision, advantages that have been highlighted in previous work~\cite{pappalardo2024photonic, facelli2024exact, wierichs2022general}. The proposed method is highly relevant to ongoing research in photonic quantum technologies, since photon loss is an unavoidable feature of current systems. A derivative computation scheme that remains reliable under these conditions could directly benefit ongoing efforts in on-hardware optimization. 

	This manuscript has been written as  follows: In Sec.~\ref{sec:theory} we provide a mostly self-contained introduction to the formalism of quantum optics and parameter-shift rules for harmonic functions. In Sec.~\ref{sec:BSandGBS} we derive the main theoretical results of our work, showing how PSRs operate for Fock Boson Sampling and how they mostly fail for Gaussian Boson Sampling.
	In Sec.~\ref{sec:experiments} we describe the hardware and the simulators. A final discussion of the results and conclusions are presented in Sec.~\ref{sec:end}.

    \section{Theory}\label{sec:theory}
    
    \subsection{Formalism} 
    
    We describe an $M$-mode bosonic system using the ladder operators: to every mode $j$, we associate an annihilation operator \(a_j\) and a creation operator \(a_j^\dagger\), satisfying the canonical commutation relation
    \begin{align}
        [a_k,a_l^\dagger] = \delta_{k, l}, \quad  [a_j,a_l] = [a_j^\dagger, a_l^\dagger] =0.
    \end{align}
    The vacuum state \(\ket{0}\) is defined as the unique state satisfying
    \begin{align}
        a_j\ket{0} = 0 \quad \forall j \in \{0,1,\ldots,M-1\}.
    \end{align}
    From the vacuum state, one can construct the Fock states (or number states) by repeatedly applying the creation operator. For a single mode, and dropping temporarily the subindex in the ladder operator, one can write the \(n\)-photon Fock state as
    \begin{align}
        \ket{n} = \frac{(a^\dagger)^n}{\sqrt{n!}} \ket{0}.
    \end{align}
    These states form an orthonormal basis of the Hilbert space associated with a single bosonic mode. They are referred to as number states because they are eigenstates of the number operator, $a^\dagger a$
    with eigenvalue \(n\):
    \begin{align}
        a^\dagger a \ket{n} = n\ket{n}.
    \end{align}
    The integer \(n\) corresponds to the number of photons occupying mode \(j\). In most quantum optics experiments, one considers several bosonic modes simultaneously. The Hilbert space of the full system is therefore given by the tensor product of the Hilbert spaces associated with each mode. For an $M$-mode system, the multimode Fock states are defined as
    \begin{align}
        \ket{\vec{n}} = \ket{n_0,n_1,\dots,n_{M-1}} = \bigotimes_{j=0}^{M-1} \ket{n_j}.
    \end{align}
    The vector
    \begin{align}
        \vec{n} = (n_0,n_1,\dots,n_{M-1}),
    \end{align}
    is usually referred to as the Fock occupation number vector of the multimode state. 
    
    Another class of useful states are the coherent states since they represent the light emitted by a laser. A single-mode coherent state of amplitude \(\alpha \in \mathbb{C}\) can be generated by applying the displacement operator to the vacuum state:
    \begin{align}
        \ket{\alpha} = D(\alpha)\ket{0},
    \end{align}
    where the displacement operator is defined as
    \begin{align}
        D(\alpha) = \exp\left(\alpha a^\dagger - \alpha^* a\right).
    \end{align}
    A single-mode coherent state can be expanded in the Fock basis as~\cite{barnett2002methods}
    \begin{align}
        \ket{\alpha} = e^{-|\alpha|^2/2} \sum_{n=0}^{\infty} \frac{\alpha^n}{\sqrt{n!}} \ket{n}.
    \end{align}
    The parameter \(\alpha\) determines the mean photon number \(\bra{\alpha} a^\dagger a \ket{\alpha} = |\alpha|^2\) which readily follows from the fact that coherent states are right (left) eigenstates of the annihilation (creation) operator 
    \begin{align}
    a\ket{\alpha} = \alpha \ket{\alpha}, \quad \bra{\alpha} a^\dagger = \alpha^* \bra{\alpha}.
    \end{align}
    
    In the multimode case, a coherent state is defined by a vector of complex amplitudes
    \begin{align}
        \vec{\alpha} = (\alpha_0, \alpha_1, \dots, \alpha_{M-1})^T,
    \end{align}
    where each \(\alpha_j\) corresponds to a different bosonic mode. The multimode coherent state can be defined as the tensor product
    \begin{align}
        \ket{\vec{\alpha}} = \ket{\alpha_0, \alpha_1, ..., \alpha_{M-1}} = \bigotimes_{j=0}^{M-1} \ket{\alpha_j}.
    \end{align}
    
    Lossless interferometers are described by unitary operators that, in the Heisenberg picture, map creation operators to linear combinations of creation operators. In particular, a linear optical unitary operator \(\mathcal{U}\) induces the transformation
    \begin{align}\label{eq:heisenberg}
        \mathcal{U} a_i^\dagger \mathcal{U}^\dagger = \sum_{j=0}^{M-1} U_{ji}\, a_j^\dagger,
    \end{align}
    where \(U \) is a unitary matrix describing the mode transformation in the Heisenberg picture. For coherent states, the action of \(\mathcal{U}\) is particularly simple. Let \(\vec{\alpha} = (\alpha_0, \dots, \alpha_{M-1})\in \mathbb{C}^M\) and
    \begin{align}
\mathcal{U} \ket{\vec{\alpha}}&=\mathcal{U}
\exp\left( \left[\sum_{j=0}^{M-1} \alpha_j a_j^\dagger - \alpha^*_j a_j \right] \right)\ket{0} \\
&=\exp\left(\mathcal{U} \left[\sum_{j=0}^{M-1} \alpha_j a_j^\dagger - \alpha^*_j a_j \right] \mathcal{U}^\dagger \right) \mathcal{U} \ket{0},
\end{align}
where we have used that $\mathcal{U}e^{O}\mathcal{U}^\dagger = e^{\mathcal{U} O \mathcal{U}^\dagger}$ and that $\mathcal{U}$ is unitary.
Now we will use the fact that \(\mathcal{U}\ket{0} = \ket{0}\) and Eq.~\eqref{eq:heisenberg}
    \begin{align}
   \mathcal{U} \ket{\vec{\alpha}}    & = \exp\!\left( \sum_{j,k=0}^{M-1} \alpha_k U_{jk} a_j^\dagger - \sum_{j,k=0}^{M-1} \alpha_k^* U_{jk}^* a_j \right) \ket{0}\\
        &= \exp\!\left( \sum_{j,k=0}^{M-1} \left(U_{jk}\alpha_k\right) a_j^\dagger - \sum_{j,k=0}^{M-1} \left(U_{jk}\alpha_k\right)^* a_j \right)\ket{0}. \nonumber
    \end{align}
    Now summing over all modes, we define the transformed amplitude vector
    \begin{align}\label{eq:trans}
      (\vec{\beta})_j =   (U\vec{\alpha})_j = \sum_{k=0}^{M-1} U_{jk}\alpha_k.
    \end{align}
    Finally,
    \begin{align}
    \label{coherent_unitary}
        \mathcal{U}\ket{\vec{\alpha}} %
        = \ket{U\vec{\alpha}} = \ket{\vec{\beta}}.
    \end{align}
Note that the action of a lossless interferometer on a product of coherent states produces another product of coherent states whose amplitudes transform according to Eq.~\eqref{eq:trans}, exactly as one would expect classically.
    
    Another useful state that we will consider is the single-mode squeezed state, which is defined from the squeezing operator with real parameter \(r\) as follows
    \begin{align}
        S(r) \ket{0}, \text{ with } S(r)  = e^{\tfrac{r}{2}(a^2 - (a^\dagger)^2)}.
    \end{align}
Using the disentangling identity~\cite{barnett2002methods} 
    \begin{align}
S(r) =         e^{-\tfrac{1}{2} \tanh(r) (a^\dagger)^2} e^{- \ln \cosh(r) \left [a^\dagger a + \tfrac{1}{2}  \right]} e^{\tfrac{1}{2} \tanh(r) a^2} 
    \end{align}
 one can write a single-mode squeezed state as
    \begin{align}
      S(r)\ket{0} =   \frac{1}{\sqrt{\cosh(r)}} e^{-\tfrac{1}{2} \tanh(r) (a^\dagger)^2} \ket{0}.
    \end{align}
    To simplify the notation, we will write a tensor product of squeezed states in $M$ modes as
    \begin{align}
        \ket{\vec{r}} := \bigotimes_{j=0}^{M-1} \ket{r_j} = \prod_{j=0}^{M-1} \frac{1}{\sqrt{\cosh(r_j)}} e^{-\tfrac{1}{2} \tanh(r_j) (a_j^\dagger)^2} \ket{0},
    \end{align}
and using previous results the action of a passive unitary on $\ket{\vec{r}}$ can be written as
    \begin{align}
        \mathcal{U} \ket{\vec{r}} &= \mathcal{N} e^{-\tfrac{1}{2} \sum_k \tanh(r_k) (\mathcal{U} a_k^\dagger \mathcal{U}^\dagger)^2} \ket{{0}} \nonumber %
        \\ &= \mathcal{N} e^{-\tfrac{1}{2} \sum_{j,l=0}^{M-1} A_{j,l}a_j^\dagger a_l^\dagger } \ket{{0}},
    \end{align}
    where the normalization constant 
    \begin{align}
    \mathcal{N}=  \frac{1}{\sqrt{ \prod_{k=0}^{M-1} \cosh(r_k)}}
    \end{align} and the symmetric matrix $A=A^T$ has Autonne-Takagi decomposition~\cite{houde2024matrix} 
    \begin{align}\label{eq:pure}
    A = U D U^T
    \text{ with  }D = \oplus_{j=0}^{M-1} \tanh r_j.
\end{align}
    \subsection{Optical Components} Although unitary evolution fully describes the ideal dynamics of a lossless bosonic system, optical experiments inevitably involve loss. Unlike unitary evolution, losses are non-unitary and cannot be directly modelled within the linear optics formalism just introduced. A standard approach is to add environmental modes to the system, into which information leaks, and then to trace them out in order to model this loss. Within this picture we can naturally accommodate sub-unitary\footnote{We define a sub-unitary matrix as a square matrix with singular values bounded by 1.} matrices $T$ that play the analogous role of the matrix $U$ that transform input-output coherent amplitudes in Eq.~\eqref{eq:trans}.
     In this larger picture, the overall evolution remains unitary, but is defined on the combined system-environment space. Specifically, the lossy evolution of an \(M\)-mode system is embedded in a larger unitary transformation \( W \) acting on \(2M\) modes~\cite{bulmer2022threshold}. The matrix \( W \) takes the form
    
    \begin{align} \label{church}
        W& = \begin{pmatrix}
        T & \sqrt{F} \\
        \sqrt{E} & -T^\dagger
        \end{pmatrix},\
    E = \id - T^\dagger T,   \ F = \id - T T^\dagger,
\end{align}
    where \(T\) is an $M\times M$ sub-unitary matrix describing the lossy transformation on the system modes and we use $\id$ for the $M \times M$ identity matrix. The additional \(M\) modes (corresponding to the indices $M,M+1,\ldots,2M-1$), corresponding to the environment labeled as the $B$ subsystem, are initialized in the vacuum state, and the full evolution remains unitary on the enlarged Hilbert space. It can be seen that \(W\) is unitary by writing the singular value decomposition of \(T\)
    \begin{align}
        T = U \Upsilon V^\dagger, \quad U,V \text{ unitary} ; \, \Upsilon_{i,j} = \Upsilon_{i,i} \delta_{i,j},
    \end{align}
    with \(0 \leq \Upsilon_{i,i} \leq 1, \, \forall i\in \{0, 1, \dots, M-1\}\). Using Eq.~\eqref{church} and denoting by $\zm$ the $M \times M$ zero matrix we have
    \begin{align}
        W = \begin{pmatrix}
            U & \zm \\ \zm & V
        \end{pmatrix} \begin{pmatrix}
        \Upsilon & \sqrt{\id - \Upsilon^2} \\ \sqrt{\id -\Upsilon^2} & -\Upsilon
        \end{pmatrix} \begin{pmatrix}
            V^\dagger & \zm \\ \zm & U^\dagger
        \end{pmatrix},
    \end{align}
    which is unitary. Thus, the evolution of the system under loss can be written as
    \begin{align}
    \label{loss_channel}
        \mathrm{Tr}_B \Big( \mathcal{W} (\rho \otimes \ket{\vec{0}}\bra{\vec{0}}) \mathcal{W}^\dagger \Big) := \mathcal{L}_T ( \rho ).
    \end{align}
where the partial trace in Eq.~\eqref{loss_channel} is performed precisely over the environmental modes. 
The infinite-dimensional Hilbert-space unitary $\mathcal{W}$ which is constructed from the mode unitary matrix $W$ acts on the system and the environment and implements the following Heisenberg picture transformation on the creation operations of the system
\begin{align}
\mathcal{W} a_i^\dagger \mathcal{W}^\dagger = \sum_{j=0}^{M-1} \left[ T_{ji}\, a_j^\dagger + (\sqrt{E})_{ji}\, b_j^\dagger \right],
\end{align}
where $b_j^\dagger$ is a creation operator for the $j$-th environmental mode.
We thus model a realistic boson sampling experiment where a state $\rho$ is prepared, a lossy linear optical transformation $T$ is applied, and photon-number measurements are performed at each output mode.

    Note that if the input state \(\rho\) is a pure Fock state, we refer to the experiment as Fock Boson Sampling. If instead \(\rho\) is a Gaussian state, such as a product of single-mode squeezed states, we refer to the experiment as Gaussian Boson Sampling (GBS).

\begin{figure}[!t]
        \centering
        \includegraphics[width=0.99\linewidth]{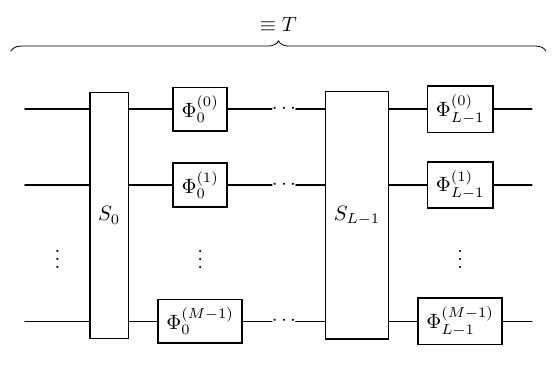}
        \caption{A lossy interferometer decomposed as a sequence of $L$ unparameterized sub-unitary blocks and parametrized diagonal phase-shift matrices.}
        \label{fig:decomposition_circuit} 
    \end{figure}

    To achieve practical quantum computing, one must design a tunable optical system capable of implementing arbitrary linear optical transformations. Such devices are commonly referred to as universal multiport interferometers. Multiple decompositions of universal interferometers have been proposed. First, it has been shown that any \(M\)-mode unitary transformation can be decomposed into a network of only two elementary optical components: 50:50 beam splitters and phase-shifters. This decomposition is realized by the architectures introduced by Clements \textit{et al.}~\cite{clements2016optimal} or Bell \textit{et al.}~\cite{bell2021further}. Alternatively, arbitrary unitary transformations can be decomposed using only discrete Fourier transform (DFT) blocks separated by rows of phase-shifters, as described by L\'opez-Pastor \textit{et al.}~\cite{pastor2021arbitrary} and Girouard \textit{et al.}~\cite{girouard2026nearoptimal}. Since losses are independent of the phase-shifters, any universal interferometer decomposition can be seen as a sequence of fixed, unparameterized sub-unitary blocks separated by diagonal phase-shift matrices. The sub-unitary blocks consist of either collections of lossy beam splitters or lossy approximations of DFT blocks, depending on the chosen decomposition. As a result, all tunable parameters of the interferometer are contained exclusively in the diagonal phase-shift matrices. Writing the fixed sub-unitary blocks for layer $i$ as $S_i$ and the diagonal phase-shift matrices for layer $i$ and mode $j$ as \(\Phi_i^{(j)} \), any such decomposition can be represented schematically as shown in Fig.~\ref{fig:decomposition_circuit}, where each \(\Phi_i^{(j)}\) is a diagonal phase-shift matrix containing the tunable parameters, while each \(S_i\) is a fixed sub-unitary transformation independent of those parameters. 
    Note that for the decompositions of Clements \textit{et al.} and Bell \textit{et al.}, the matrices \(S_i\) are highly sparse, as each corresponds to an array of beam splitters that couples only adjacent modes. 
    
    To verify that the assumptions made on the structure of the sub-unitary matrices are physically meaningful, we examine how such optical components are implemented in actual systems. In practice, beam-splitters are made using evanescently coupled waveguides in integrated photonic circuits, which are lithographically designed to achieve a target splitting ratio of 50:50~\cite{clements2016optimal, bombardelli2025foundations, dong2023programmable}. Phase-shifters are made by locally modifying the effective length or refractive index of a waveguide, most commonly using thermo-optic heaters or via electro-optic effects respectively~\cite{clements2016optimal, bombardelli2025foundations, dong2023programmable, reed2010silicon, poveda2025etching}. Since the optical path modification as a result of the phase-shift is at most on the order of the wavelength of the light, one can assume without loss of generality that the phase-shifter does not significantly impact the transmission~\cite{poveda2025etching, psiquantum2025manufacturable}. Consequently, we assume that loss happening around a phase-shifter $i,j$ commutes with the action of the phase-shifter when cross-talk between different modes is negligible, and thus each single mode operator in Fig.~\ref{fig:decomposition_circuit} is of the form 
    \begin{align}\label{eq:phase}
    \Phi_i^{(j)} = \sqrt{\eta_{i,j}} e^{i \theta_{i,j}}, 
    \end{align}
    where $\eta$ is an energy transmission factor and $\theta$ is a tunable parameter. In the following section we will develop parameter-shift rules to obtain the gradient of photon-number probabilities precisely with respect to these phases.

    Finally, photons are measured using single-photon detectors positioned at the output modes of the interferometer. Depending on the application, either photon-number-resolving detectors or threshold detectors may be used. Threshold detectors cannot distinguish the exact number of detected photons, but only whether no photons or at least one photon was detected. Common detector technologies include avalanche photodiodes, superconducting nanowire single-photon detectors, and transition-edge sensors, each offering different trade-offs in detection efficiency, timing resolution, and dark-count rate~\cite{hadfield2009single,eisaman2011invited,dalbec2025accurate}.

    \subsection{Parameter-Shift Rules}
    In this subsection we discuss parameter-shift rules mostly following the presentation from Ref.~\cite{pappalardo2024photonic}. Consider an arbitrary observable quantity associated with a photonic circuit, such as a measurement outcome probability, an average photon number or a photon-number correlator. If the dependence of such an observable on a given phase-shifter parameter \(\theta\) can be expressed as
    \begin{align}
    \label{eq_PSR}
        f(\theta) = \sum_{m=-n}^{n} k_m e^{im\theta}, \, k_m \in \mathbb{C}
    \end{align}
    for some finite \(n\) then one can obtain parameter-shift rules to accurately compute its derivative with respect to $\theta$.  Note that most useful linear optical observables can be written in this way, as their are functions of the observed photon-number probabilities.  To this end, one can differentiate with respect to \(\theta\) to obtain
    \begin{align}
    \label{eq_PSR2}
        \partial_\theta f(\theta) = i\sum_{m=-n}^{n} m k_m e^{im\theta}.
    \end{align}
    The expression for the derivative in Eq.~\eqref{eq_PSR2} has the same finite \(\theta\) dependency as in Eq.~\eqref{eq_PSR}. Consequently, the derivative can be reconstructed from a suitable linear combination of shifted evaluations of \(f\). This observation forms the basis of PSRs~\cite{pappalardo2024photonic,facelli2024exact,wierichs2022general}. Let \(\{\mu_l\}\) denote a set of phase-shifts and \(\{c_l\}\) the corresponding coefficients. A PSR takes the form
    \begin{align}
    \label{eq_PSR3}
        \partial_\theta f(\theta) = \sum_{l=-n}^{n} c_l\,f(\theta+\mu_l).
    \end{align}
    Using Eq.~\eqref{eq_PSR} and Eq.~\eqref{eq_PSR2}, one can write Eq.~\eqref{eq_PSR3} as
    \begin{align}
        i \sum_{m = -n}^{n} m k_m e^{i m \theta} = \sum_{l = -n}^{n} c_l \sum_{m = -n}^{n} k_m e^{i m (\theta+\mu_l)}.
    \end{align}
    Since the expressions on both sides are polynomials, we can group the powers together and get
    \begin{align}
        imk_m e^{im\theta} = k_m e^{im\theta} \sum_{l = -n}^{n} c_l e^{im\mu_l}, 
    \end{align}
    which for \(k_m \neq 0\) gives the system of equations
    \begin{align}
        im = \sum_{l = -n}^{n} c_l e^{im\mu_l} \quad \forall m \in [ -n, n ].
    \end{align}
    We can formally write the previous expression as a system of equations where we will arbitrarily order the indices as such \((0, 1, \dots, n, -n, \dots, -2, -1)\), we get
    \begin{align}
    \label{lin_system}
        \begin{pmatrix}
            1 & 1 & \dots & 1 \\ 
            e^{i\mu_0} & e^{i\mu_1} & \dots & e^{-i\mu_{-1}} \\ \vdots & \vdots & \ddots & \vdots \\ e^{-i\mu_0} & e^{-i\mu_1} & \dots & e^{i\mu_{-1}}
        \end{pmatrix} \cdot \begin{pmatrix}
            c_0 \\ c_1 \\ \vdots \\ c_{-1}
        \end{pmatrix} = i \begin{pmatrix}
            0 \\ 1 \\ \vdots \\ -1
        \end{pmatrix}.
    \end{align}
    Thus if this system has a solution for a given array of angles \(\{\mu_l\}\), then one can reconstruct the derivative by solving the linear system for the constants \(\{c_l\}\). Notice that from Eq.~\eqref{eq_PSR2} one can deduce that the coefficient of the constant term is 0. Using the same logic, one can notice that the constant parts cancel out by symmetry i.e. \(c_{-l} = -c_l\), which yields
    \begin{align}
        \begin{pmatrix}
            1 & 1 & \dots & 1 \\ 
            e^{i\mu_0} & e^{i\mu_1} & \dots & e^{-i\mu_{-1}} \\ \vdots & \vdots & \ddots & \vdots \\ e^{-i\mu_0} & e^{-i\mu_1} & \dots & e^{i\mu_{-1}}
        \end{pmatrix} \cdot \begin{pmatrix}
            0 \\ c_1 \\ \vdots \\ -c_{1}
        \end{pmatrix} = i \begin{pmatrix}
            0 \\ 1 \\ \vdots \\ -1
        \end{pmatrix}.
    \end{align}
    The equations we get for this system are of the form
    \begin{align}
        im = \sum_{l=1}^{n} c_l \Big( e^{im\mu_l} - e^{im\mu_{-l}} \Big),
    \end{align}
    which can be simplified to 
    \begin{align}
        m = \sum_{l=1}^{n} 2 c_l e^{im(\mu_l + \mu_{-l})/2} \sin\Big(\tfrac{m(\mu_l - \mu_{-l})}{2}\Big).
    \end{align}
    We can add the equations corresponding to \(m\) and \(-m\) to get
    \begin{align}
        0 &= \sum_{l=1}^{n} c_l \sin\Big(\tfrac{m(\mu_l + \mu_{-l})}{2}\Big) \sin\Big(\tfrac{m(\mu_l - \mu_{-l})}{2}\Big) \quad \forall m \nonumber \\ &\implies \cos(m\mu_l) = \cos(m\mu_{-l}) \quad \forall l, m \in [-n, n].
    \end{align}
    Since letting \(\mu_l = \mu_{-l}\) makes the matrix singular, then we must set \(\mu_{-l} = -\mu_l\). Replacing this into the system of equations yields
    \begin{align}
        \frac{m}{2} = \sum_{l=1}^{n} c_l \sin(m \mu_l) \quad \forall m > 0,
    \end{align}
    which corresponds to the system
    \begin{align*}
        \begin{pmatrix}
            \sin(\mu_1) & \sin(\mu_2) & \dots & \sin(\mu_n) \\
            \sin(2\mu_1) & \sin(2\mu_2) & \dots & \sin(2\mu_n) \\
            \vdots & \vdots & \ddots & \vdots \\
            \sin(n\mu_1) & \sin(n\mu_2) & \dots & \sin(n\mu_n)
        \end{pmatrix}
        \cdot \begin{pmatrix}
            c_1 \\
            c_2 \\
            \vdots \\
            c_n
        \end{pmatrix}
        =
        \frac{1}{2}
        \begin{pmatrix}
            1 \\
            2 \\
            \vdots \\
            n
        \end{pmatrix}.
    \end{align*}
    Consequently, any set of shifts for which the associated matrix is invertible defines a valid PSR, and since both the matrix and the target vector are purely real, the solution coefficients \(c_l\) are also purely real. Note that by letting \(\mu_m = \tfrac{2\pi m}{2n + 1}\), we get the result found in~\cite{pappalardo2024photonic} which makes the matrix in Eq.~\eqref{lin_system} proportional to the DFT matrix,  giving the best possible choice of angles for noisy systems, as it maximizes the spacing between shifts, increasing error tolerance. %

Besides PSRs one can estimate gradients using other methods. A common one is the Finite Difference method, which estimates the derivative of a target function by evaluating it at two nearby parameter values. It relies on the approximation 
    \begin{align}
        f'(\theta) \approx \frac{f(\theta+\delta)-f(\theta-\delta)}{2\delta},
    \end{align}
    where \(\delta\) is a small parameter-shift. This approximation is obtained from a second-order Taylor expansion and has a truncation error of order \(\mathcal{O}(\delta^2)\)~\cite{teo2023optimized, liang2025qugstep}. In practice, the choice of \(\delta\) involves a trade-off: large values of \(\delta\) increase the approximation error, whereas very small values can amplify statistical fluctuations in noisy systems through the subtraction of two nearly identical function evaluations. Consequently, the performance of the FD method is highly dependent on the choice of \(\delta\).

\section{Parameter-Shift Rules in Fock and Gaussian Boson Sampling}\label{sec:BSandGBS}
    \subsection{Fock State Transition Probability} For a pure Fock state, the transition probability from an initial state \(\vec{i} = (i_0,\ldots,i_{M-1})\) to a final state \(\vec{j}=(j_0,\ldots,j_{M-1})\) can be written using Eq.~\eqref{loss_channel}
    \begin{align}
    \label{transitionprob}
   \braket{\vec{j}| \mathcal{L}_T(\ket{i} \bra{i})|\vec{j}} =   \bra{\vec{j}} \mathrm{Tr}_B \Big( \mathcal{W} \ket{\vec{i}, \vec{0}} \bra{\vec{i}, \vec{0}} \mathcal{W}^\dagger \Big) \ket{\vec{j}}. %
    \end{align}
    which is difficult to evaluate directly. For a multimode coherent state, one can use Eq.~\eqref{coherent_unitary} and Eq.~\eqref{loss_channel} to write
    \begin{align}
        \mathrm{Tr}_B \Big( \mathcal{W} \ket{\vec{\alpha}, \vec{0}} \bra{\vec{\alpha}, \vec{0}} \mathcal{W}^\dagger \Big) &= \ket{T\vec{\alpha}} \bra{T\vec{\alpha}},
    \end{align}
    which is a much simpler expression than Eq.~\eqref{transitionprob} and follows from the fact that coherent states do not get entangled in interferometers. 
    We can use this property of coherent states to find an equivalent expression for Eq.~\eqref{loss_channel}. For two coherent states \( \ket{\vec{\alpha}} \) and \( \ket{\vec{\beta}} \), we can easily obtain the following equality
    \begin{align}
        \bra{\vec{\beta}} \mathcal{L}_T \left( \ket{\vec{\alpha}}\bra{\vec{\alpha}} \right) \ket{\vec{\beta}} &= \nonumber \\ \bra{\vec{\beta}} e^{-\| \vec{\alpha} \|^2} \mathcal{L}_T \Bigg( \sum_{\vec{i}, \vec{j} \geq \vec{0}} &\frac{(\vec{\alpha})^{\vec{i}} (\alpha^*)^{\vec{j}}}{\sqrt{\vec{i}! \vec{j}}!} \ket{\vec{i}}\bra{\vec{j}} \Bigg) \ket{\vec{\beta}}.
    \end{align}
    Using linearity and expanding the remaining coherent states we obtain
    \begin{align}
        = e^{-\|\vec{\alpha}\|^2 - \| \vec{\beta}\|^2} \sum_{\vec{i}, \vec{j}, \vec{k}, \vec{l} \geq \vec{0}} \frac{(\vec{\alpha})^{\vec{i}} (\vec{\alpha}^*)^{\vec{j}} (\vec{\beta})^{\vec{k}} (\vec{\beta}^*)^{\vec{l}}}{\sqrt{\vec{i}! \vec{j}! \vec{k}! \vec{l}!}} \nonumber \\ \cdot \bra{\vec{l}} \mathcal{L}_T \left( \ket{\vec{i}}\bra{\vec{j}} \right) \ket{\vec{k}},
    \end{align}
where we use the following notation $\vec{i}! = \prod_{j=0}^{M-1} (i_{j})!$, $\vec{\alpha}^{\vec i} = \prod_{j=0}^{M-1} \alpha_j^{i_j}$.
    Notice that the expression can also be expanded using the action of the sub-unitary \( T \) on the coherent states. We obtain
    \begin{align}
        \braket{\vec{\beta} | T\vec{\alpha}} \braket{T\vec{\alpha} | \vec{\beta}} 
        = e^{ -\|\vec{\beta}\|^2 - \|T\vec{\alpha}\|^2 + \vec{\beta}^\dagger T \vec{\alpha} + \vec{\alpha}^\dagger T^\dagger \vec{\beta} },
    \end{align}
where we use the fact that the overlap of two multimode coherent states is $|\braket{\vec{\alpha}|\vec{\beta}}|^2 = \exp(-||\vec{\alpha}-\vec{\beta}||^2)$.
    Combining the two expressions yields
    \begin{align}
        e^{\|\vec{\alpha}\|^2 - \|T\vec{\alpha}\|^2 + \vec{\beta}^\dagger T \vec{\alpha} + \vec{\alpha}^\dagger T^\dagger \vec{\beta}} &= \nonumber \\ \sum_{\vec{i}, \vec{j}, \vec{k}, \vec{l} \geq \vec{0}} \frac{(\vec{\alpha})^{\vec{i}} (\alpha^*)^{\vec{j}} (\vec{\beta})^{\vec{k}} (\beta^*)^{\vec{l}}}{\sqrt{\vec{i}! \vec{j}! \vec{k}! \vec{l}}} &\bra{\vec{l}} \mathcal{L}_T \left( \ket{\vec{i}}\bra{\vec{j}} \right) \ket{\vec{k}}.
    \end{align}
    We can take the derivative at \(\vec{\alpha} = \vec{\beta} = \vec{\alpha}^* = \vec{\beta}^* =  \vec{0}\) on both sides. We obtain
    \begin{align}
        \left. \partial_{\vec{\alpha}}^{\vec{i}} \, 
        \partial_{\vec{\alpha}^*}^{\vec{j}} \, 
        \partial_{\vec{\beta}}^{\vec{k}} \,
        \partial_{\vec{\beta}^*}^{\vec{l}} 
        e^{\|\vec{\alpha}\|^2 - \|T\vec{\alpha}\|^2 + \vec{\beta}^\dagger T \vec{\alpha} + \vec{\alpha}^\dagger T^\dagger \vec{\beta}} 
        \right|_{\vec{\alpha} = \vec{\beta} = 0}
        = \nonumber \\ \sqrt{\vec{i}! \, \vec{j}! \, \vec{k}! \, \vec{l}!} \ 
        \bra{\vec{l}} \mathcal{L}_T \big( \ket{\vec{i}}\bra{\vec{j}} \big) \ket{\vec{k}}.
    \end{align}
   To simplify the notation, let \begin{align}
        \vec{z} = \begin{pmatrix}
        \vec{\alpha} \\ \vec{\alpha}^* \\ \vec{\beta} \\ \vec{\beta}^*
    \end{pmatrix} ; \quad \vec{\omega} = \begin{pmatrix}
        \vec{i} \\ \vec{j} \\ \vec{k} \\ \vec{l}
    \end{pmatrix},
    \end{align}
    and define the matrix
    \begin{align}
        \Sigma = \begin{pmatrix}
            \zm & E^* & \zm & T^T \\ E & \zm & T^\dagger & \zm \\ \zm & T^* & \zm & \zm \\ T & \zm & \zm & \zm
        \end{pmatrix},
    \end{align}
    so that we can write
 \begin{align}
        \vec{\alpha}^\dagger E \vec{\alpha} + \vec{\beta}^\dagger T \vec{\alpha} + \vec{\alpha}^\dagger T^\dagger \vec{\beta} = \tfrac12 \vec{z}^T \Sigma z
\end{align}
    where we can recall that $E = \id - T^\dagger T $ (cf. Eq.~\eqref{church}). We can introduce~\cite{yao2024riemannian, banchi2020training} 
    \begin{align}
    \label{hafnian}
        \partial_{\vec{z}}^{\vec{\omega}} e^{\frac{1}{2} \vec{z}^T \Sigma \vec{z}}|_{\vec{z} = \vec{0}} = \text{Haf} \left[ \Sigma_{\vec{
        \omega}} \right],
    \end{align}
    where \(\Sigma_{\vec{\omega}}\) is obtained by repeating the rows and columns of \(\Sigma\) according to the vector \(\vec{\omega}\), that is, column and row $i$ of $\Sigma$ are repeated a total of $\omega_i$ times.    
    If $\omega_i=0$ column and row $i$ are removed. Note that $\Sigma_{\vec{\omega}}$ is of even size $2n = \sum_i \omega_i$.
    Finally, the hafnian of a matrix $A$ of size $2n \times 2n$ is given by
    \begin{align}
    \mathrm{Haf} [A ] := \sum_{\sigma \in P_{2n}} \prod_{\{\mu, \nu\} \in \sigma} A_{\mu, \nu}. 
\end{align}
The set $P_{2n}$ contains all the possible pairings of \(2\) elements in $\{0,1,\ldots, 2n-1\}$, also known as the set of perfect-matching permutations.
    
    Notice that any permutation can be applied to the matrix $\Sigma$, provided we apply this transformation to the vector of indices \(\vec{\omega}\), since
    \begin{align}
        \vec{z}^T \Sigma \vec{z} = \vec{z}^T P^T (P \Sigma P^T) P \vec{z}.
    \end{align}
    Using the permutation 
    \begin{align}
        P_{0, 2, 1, 3} = \begin{pmatrix}
            \id & \zm & \zm & \zm \\ \zm & \zm & \id & \zm \\ \zm & \id & \zm & \zm \\ \zm & \zm & \zm & \id 
        \end{pmatrix}, 
    \end{align} 
    and plugging into \eqref{hafnian} yields
    \begin{align*}
        \bra{\vec{l}} \mathcal{L}_T \left( \ket{\vec{i}}\bra{\vec{j}} \right) \ket{\vec{k}} = \frac{1}{\sqrt{\vec{i}! \vec{j}! \vec{k}! \vec{l}!}} \text{Haf} \left[\begin{pmatrix}
        \zm_{2M} & B^T \\ B & \zm_{2M}
        \end{pmatrix}_{\vec{i} \oplus \vec{k} \oplus \vec{j} \oplus \vec{l}} \right],
    \end{align*}
    where
    \begin{align}\label{eq:bdef}
        B = \begin{pmatrix}
        E & T^\dagger \\ T & \zm
        \end{pmatrix},
    \end{align}
    and $\vec{a} \oplus \vec{b}$ is the concatenation of the vectors $\vec a$ and $\vec b$. This gives us an expression for the transition probability
    \begin{align}
        \bra{\vec{j}} \mathcal{L}_T \left( \ket{\vec{i}}\bra{\vec{i}} \right) \ket{\vec{j}} &= \frac{1}{\vec{i}! \vec{j}!} \text{Haf} \left[ \begin{pmatrix}
        \zm & B \\ B^T & \zm
        \end{pmatrix}_{\vec{i} \oplus \vec{j} \oplus \vec{i} \oplus \vec{j}} \right],\\
 &  = \label{perm_prob}
         \frac{1}{\vec{i}! \vec{j}!} \mathrm{Perm} \left[
        \left(
        \begin{array}{cc}
        \id - T^\dagger T & T^\dagger \\
        T & \zm
        \end{array}
        \right)_{\vec{i} \oplus \vec{j}}
        \right],
    \end{align}
by using the identity
\begin{align}
\text{Haf}\left[ \begin{pmatrix}
    \zm & B \\ B^T & \zm
        \end{pmatrix} \right] = 
\text{Haf} \left[\begin{pmatrix}
        \zm & B^T \\ B & \zm
        \end{pmatrix} \right]= \text{Perm}(B). 
\end{align}
For an $n \times n$ matrix $B$ we write its permanent as 
    \begin{align}
\text{Perm}(B) =\sum_{\sigma \in S_n} \prod_{i=0}^{n-1} B_{i,\sigma(i)}
    \end{align}
    where $S_n$ is the symmetric group action on $n$ objects. Note the similarity of the equation above with the definition of the determinant. Note that Eq~\eqref{perm_prob} coincides with the result from Ref.~\cite{yao2024riemannian}. However, our result is derived without resorting to the Choi–Jamio{\l}kowski isomorphism.

We now want to understand how the transmission depends on the phase-shift parameters. Consider the phase at position $(k,l)$ that we will generically call $\theta$. From Eq.~\eqref{eq:phase} together with Fig.~\ref{fig:decomposition_circuit}, each entry of the net transmission matrix has the following functional dependence for the particular $\theta$ we are considering
    \begin{align}
        T_{m, n} = a e^{i\theta} + b; \quad a, b \in \mathbb{C}.
    \end{align}
    Using Eq.~\eqref{eq:bdef}, we have 
    \begin{align}\label{eq:blah}
        B_{m, n} = a'e^{i\theta} + b'e^{-i\theta} + c' ; \quad a', b', c' \in \mathbb{C}.
    \end{align}
    Note that $\sum_{k=0}^{M-1} j_k \leq \sum_{k=0}^{M-1} i_k = n$ as one cannot detect more photons at the output than photons at the input. 
    As a consequence of the three observations just mentioned, the permanent giving the probabilities is necessarily of the functional form
    \begin{align}
        \frac{1}{\vec{i}! \vec{j}!}\mathrm{Perm}\Big[B_{\vec{i} \oplus \vec{j}}\Big] = \sum_{m = -n}^{n} k_m e^{i m \theta}, \, k_m \in \mathbb{C}.
    \end{align}
    Thus, the derivative of the transition probability for Fock states can be obtained using an $n$-th order PSR, which is the main result of this subsection.
    
    One can naturally extend the result to threshold detectors ~\cite{bulmer2022threshold}. Let \(\vec{d}\) denote the output click pattern of the system. Since threshold detectors are used, \(d_i\) is a binary value, a \(0\) indicates no click and a \(1\) indicates a click in the given mode. The corresponding transition probability is defined as
    \begin{align}
        \mathrm{Tr} \Big( \mathcal{L}_T \left( \ket{\vec{i}}\bra{\vec{i}} \right) \cdot \Pi_{\vec{d}} \Big),
    \end{align}
    where
    \begin{align}
        \Pi_{\vec{d}} := \sum_{\vec{k}\in\mathcal{G}_n(\vec{d})} \ket{\vec{k}}\bra{\vec{k}}.
    \end{align}
    Here, \(\mathcal{G}_n(\vec{d})\) denotes the set of all Fock states consistent with the observed click pattern \(\vec{d}\) and containing at most \(n\) photons. Consequently,
    \begin{align*}
        \mathrm{Tr} \Big( \mathcal{L}_T \left( \ket{\vec{i}}\bra{\vec{i}} \right) \cdot \Pi_{\vec{d}} \Big) = \sum_{\vec{k}\in\mathcal{G}(\vec{d},n)} \bra{\vec{k}} \mathcal{L}_T \left( \ket{\vec{i}}\bra{\vec{i}} \right) \ket{\vec{k}}.
    \end{align*}
    Since each threshold detector transition probability is a sum of Fock transition probabilities, it follows immediately that it satisfies the same PSR with parameter \(n=|\vec{i}|\). %

    \subsection{Gaussian State Transition Probability} 
    In GBS, the probability of measuring $\vec{j}$ photons at the output when a product of single-mode squeezed states are sent into an interferometer with transmission matrix $T$ is given by
    \begin{align}
        \bra{\vec{j}} \mathcal{L}_T \left( \ket{\vec{r}}\bra{\vec{r}} \right) \ket{\vec{j}}.
    \end{align}
    As before, we will first evaluate the probability associated with coherent states\begin{align}
        e^{\|\vec{\alpha}\|^2} \bra{\vec{\alpha}} \mathcal{L}_T \left( \ket{\vec{r}}\bra{\vec{r}} \right) \ket{\vec{\alpha}} &= \nonumber \\ \sum_{\vec{k}, \vec{l} \geq \vec{0}} \frac{(\vec{\alpha})^{\vec{k}}  (\vec{\alpha}^*)^{\vec{l}}}{\sqrt{\vec{k}!\vec{l}!}} &\bra{\vec{l}} \mathcal{L}_T \left( \ket{\vec{r}}\bra{\vec{r}} \right) \ket{\vec{k}}.
    \end{align}
    Recall how we defined the loss channel in Eq.~\eqref{loss_channel}. We can use the definition and write
    \begin{align*}
        \bra{\vec{\alpha}} \mathcal{L}_T \left( \ket{\vec{r}}\bra{\vec{r}} \right) \ket{\vec{\alpha}} = \bra{\vec{\alpha}}\mathrm{Tr}_B \Big( \mathcal{W} \ket{\vec{r}, \vec{0}} \bra{\vec{r}, \vec{0}} \mathcal{W}^\dagger \Big) \ket{\vec{\alpha}}. 
    \end{align*}
    We can recall that one can write the $M$-mode identity operator over the environment modes as 
    \begin{align}
        \mathcal{I}_{B} = \int \frac{d^{2M} \vec{\beta}}{\pi^M} \ket{\vec{\beta}} \bra{\vec{\beta}},
    \end{align}
    to write
    \begin{align}
        \bra{\vec{\alpha}} &\mathcal{L}_T \left( \ket{\vec{r}}\bra{\vec{r}} \right) \ket{\vec{\alpha}} = \bra{\vec{\alpha}}\mathrm{Tr}_B \Big( \mathcal{W} \ket{\vec{r}, \vec{0}} \bra{\vec{r}, \vec{0}} \mathcal{W}^\dagger \nonumber \\ &\cdot \mathcal{I}_A \otimes \int \frac{d^{2M} \vec{\beta}}{\pi^M} \ket{\vec{\beta}} \bra{\vec{\beta}} \Big) \ket{\vec{\alpha}} \nonumber \\ &= \int \frac{d^{2M} \vec{\beta}}{\pi^M} \mathrm{Tr} \Big( \mathcal{W} \ket{\vec{r}, \vec{0}} \bra{\vec{r}, \vec{0}} \mathcal{W}^\dagger \ket{\vec{\alpha}, \vec{\beta}} \bra{\vec{\alpha}, \vec{\beta}} \Big) \nonumber \\ &= \int \frac{d^{2M} \vec{\beta}}{\pi^M} \bra{\vec{\alpha}, \vec{\beta}} \mathcal{W} \ket{\vec{r}, \vec{0}} \bra{\vec{r}, \vec{0}} \mathcal{W}^\dagger \ket{\vec{\alpha}, \vec{\beta}}.
    \end{align}
Recall that the overlap has the form
\begin{align}
    \bra{\vec{\alpha}, \vec{\beta}} \mathcal{W} \ket{\vec{r}, \vec{0}}%
    &= \mathcal{N} e^{-\tfrac{1}{2} \vec{v}^\dagger W D' W^T \vec{v}^*} \braket{\vec{\alpha}, \vec{\beta} | \vec{0}, \vec{0}} \nonumber \\ &=\mathcal{N} e^{-\tfrac{1}{2} \vec{v}^\dagger W D' W^T \vec{v}^* -\tfrac{1}{2} (\| \vec{\alpha} \|^2 + \| \vec{\beta} \|^2)},
\end{align}
where 
\begin{align}
D' = D \oplus \zm , \quad D = \oplus_{j=0}^{M-1} \tanh r_i, \\
\mathcal{N} =\frac{1}{\sqrt{\prod_{i=0}^{M-1} \cosh r_i}} \text{ and }\vec{v} = \begin{pmatrix}
    \vec{\alpha} \\ \vec{\beta}
\end{pmatrix}
\end{align} and we used the fact that coherent states are left eigenstates of the creation operator. Taking the norm squared, we get
\begin{align}
    &e^{\|\vec{\alpha}\|^2}\bra{\vec{\alpha}} \mathcal{L}_T \left( \ket{\vec{r}}\bra{\vec{r}} \right) \ket{\vec{\alpha}} = \nonumber \\ \int \frac{d^{2M} \vec{\beta}}{\pi^M}\mathcal{N}^2 &\cdot e^{-\tfrac{1}{2} \vec{v}^T W^* D' W^\dagger \vec{v}-\tfrac{1}{2} \vec{v}^\dagger W D' W^T \vec{v}^*- \| \vec{\beta} \|^2}.
\end{align}
Notice that if we define
\begin{align*}
    \vec{z} := \begin{pmatrix}
        \vec{\alpha} \\ \vec{\alpha}^* \\ \vec{\beta} \\ \vec{\beta}^*
    \end{pmatrix} = \begin{pmatrix}
        \vec{x} \\ \vec{y}
    \end{pmatrix}; \quad A':= W D' W^T = \begin{pmatrix}
        A_{00} & A_{01} \\ A_{10} & A_{11}
    \end{pmatrix},
\end{align*}
then we can write
\begin{align*}
    e^{\|\vec{\alpha}\|^2}\bra{\vec{\alpha}} \mathcal{L}_T \left( \ket{\vec{r}}\bra{\vec{r}} \right) \ket{\vec{\alpha}} = \mathcal{N}^2 \int \frac{d^{2M} \vec{\beta}}{\pi^M} e^{-\tfrac{1}{2} \vec{z}^T \Lambda \vec{z}}, 
\end{align*}
with 
\begin{align}
    \Lambda = \begin{pmatrix}
        A_{00}^\dagger & \zm & A_{01}^\dagger & \zm \\ \zm & A_{00} & \zm & A_{01} \\ A_{10}^\dagger & \zm & A_{11}^\dagger & \id \\ \zm & A_{10} & \id & A_{11} 
    \end{pmatrix}. 
\end{align}
Notice that
\begin{align}
    \vec{z}^\dagger = \vec{z}^T \cdot P_{1, 0, 3, 2} \Longleftrightarrow \vec{z}^T = \vec{z}^\dagger \cdot P_{1, 0, 3, 2} 
\end{align}
with
\begin{align}
    P_{1, 0, 3, 2} = \begin{pmatrix}
        \zm & \id & \zm & \zm \\ \id & \zm & \zm & \zm \\ \zm & \zm & \zm & \id \\ \zm & \zm & \id & \zm
    \end{pmatrix}.
\end{align}
This specific permutation by block will be denoted as \(P\) for simplicity. We get
\begin{align}
    e^{\|\vec{\alpha}\|^2} \bra{\vec{\alpha}} \mathcal{L}_T \left( \ket{\vec{r}}\bra{\vec{r}} \right) \ket{\vec{\alpha}} = \mathcal{N}^2 \int \frac{d^{2M} \vec{\beta}}{\pi^M} e^{-\tfrac{1}{2} \vec{z}^\dagger P \Lambda \vec{z}}.  
\end{align}
Also, notice that 
\begin{align*}
    S^\dagger \cdot \vec{z} = \begin{pmatrix}
        \text{Re}\{\vec{\alpha}\} \\ \text{Im}\{\vec{\alpha}\} \\ \text{Re}\{\vec{\beta}\} \\ \text{Im}\{\vec{\beta}\} 
    \end{pmatrix} = \begin{pmatrix}
        \vec{t} \\ \vec{u} 
    \end{pmatrix} = \vec{w}; \, S = \frac{1}{\sqrt{2}}\begin{pmatrix}
            R & 0 \\ 0 & R
        \end{pmatrix},
\end{align*}
where \(R = \tfrac{1}{\sqrt{2}} \begin{pmatrix}
    \id & i\id \\ \id & -i\id
\end{pmatrix}\), which allows us to write %
\begin{align}
e^{\|\vec{\alpha}\|^2} \bra{\vec{\alpha}} \mathcal{L}_T &\left( \ket{\vec{r}}\bra{\vec{r}} \right) \ket{\vec{\alpha}}=\mathcal{N}^2 \int \frac{d^{2M} \vec{u}}{\pi^M} e^{-\tfrac{1}{2} \vec{w}^T 4 S^\dagger P \Lambda S \vec{w}} \nonumber \\ &= \mathcal{N}^2\int \frac{d^{2M} \vec{u}}{\pi^M} e^{-\tfrac{1}{2} \vec{w}^T \Lambda' \vec{w}} \nonumber \\ &= \mathcal{N}^2 \frac{e^{-\tfrac{1}{2} \vec{x}^T X (\Lambda_{tt}' - \Lambda_{tu}'[\Lambda_{uu}']^{-1}\Lambda_{ut}') \vec{x}}}{ \sqrt{\mathrm{det}(\Lambda_{uu})}},
\end{align}
where we used that fact that $X = R^* R^\dagger = \left(\begin{smallmatrix} \zm & \id \\ \id  & \zm\end{smallmatrix} \right)$ and that
\begin{align}
    \tfrac{1}{\sqrt{2}}R^\dagger \cdot \vec{x} = \vec{t},
\end{align}
and introduced
\begin{align}
    \Lambda'_{tt} &= \Gamma \Omega \Gamma^\dagger, \\ \
    \Lambda'_{tu} &=  [\Lambda'_{ut}]^\dagger = \Gamma \Omega \sqrt{\Xi}, \\
    \Lambda'_{uu} &= \id \oplus \id + \sqrt{\Xi} \Omega \sqrt{\Xi}.
\end{align}
where
\begin{align}
    \Gamma = \begin{pmatrix}
        T & \zm \\ \zm & T^*
    \end{pmatrix}, \, \Xi = \begin{pmatrix}
        E & \zm \\ \zm & E^*
    \end{pmatrix}, \, \Omega = \begin{pmatrix}
        \zm & D \\ D & \zm
    \end{pmatrix}. 
\end{align}
We can define 
\begin{align}
    \Sigma &:= %
    -X(\Lambda'_{tt} - \Lambda'_{tu}[\Lambda'_{uu}]^{-1}\Lambda_{ut}).
\end{align}
To get a final expression for the \(\Sigma\) matrix, we write
\begin{align}
\label{sigma_def}
    \Sigma = -X \Gamma \Big( \Omega - \Omega \sqrt{\Xi} \Big[ \id + \sqrt{\Xi} \Omega \sqrt{\Xi} \Big]^{-1} \sqrt{\Xi} \Omega \Big) \Gamma^\dagger.
\end{align}
One can simplify Eq.~\eqref{sigma_def} as
\begin{align}
    \Sigma &= -X \Gamma \Big( \Omega - \Omega \sqrt{\Xi} \Big[ \id + \sqrt{\Xi} \Omega \sqrt{\Xi} \Big]^{-1} \sqrt{\Xi} \Omega \Big) \Gamma^\dagger, \nonumber \\ 
    &=-X \Gamma \Big( \Omega - \Omega \Big[ \Xi^{-1} + \Omega \Big]^{-1} \Omega \Big) \Gamma^\dagger, \nonumber \\ &=-X \Gamma \Big( \Omega - \Big[ \Omega^{-1} + \Omega^{-1} \Xi^{-1} \Omega^{-1} \Big]^{-1} \Big) \Gamma^\dagger.
\end{align}
Using the Woodbury identity~\cite{woodbury1950inverting}, we obtain
\begin{align}\label{eq:kernel}
    \Sigma& = -X \Gamma \Big( \Omega^{-1} + \Xi \Big)^{-1} \Gamma^\dagger \\
&= -\begin{pmatrix}
        \zm & T^* \\ T & \zm
    \end{pmatrix}\begin{pmatrix}
        \id - T^\dagger T & D^{-1} \\ D^{-1} & \id - T^T T^*
    \end{pmatrix}^{-1}\begin{pmatrix}
        T^\dagger & \zm \\ \zm & T^T
    \end{pmatrix}. \nonumber
\end{align}
Notice that in the case where $T$ is unitary then \begin{align}
\Sigma = -(A^* \oplus A),
\end{align}
where $A$ is given in Eq.~\eqref{eq:pure}. 

Finally, we obtain the following expression
\begin{align}
    \sum_{\vec{k}, \vec{l} \geq \vec{0}} \frac{(\vec{\alpha})^{\vec{k}}  (\vec{\alpha}^*)^{\vec{l}}}{\sqrt{\vec{k}!\vec{l}!}} \bra{\vec{l}} \mathcal{L}_T \left( \ket{\vec{r}}\bra{\vec{r}} \right) \ket{\vec{k}} \nonumber \\ = \frac{\mathcal{N}^2}{\sqrt{\mathrm{det}(\Lambda'_{uu})}} \exp\left(\tfrac12 \vec{x}^T \Sigma \vec{x} \right). 
\end{align}
Taking the derivative with respect to \(\vec{x}\) we obtain 
\begin{align}
    \sqrt{\vec{k}! \vec{l}!}\bra{\vec{l}} &\mathcal{L}_T \left( \ket{\vec{r}}\bra{\vec{r}} \right) \ket{\vec{k}} \nonumber \\ &= \frac{\mathcal{N}^2}{\sqrt{\mathrm{det}(\Lambda_{uu}')}} \partial_{\vec{\alpha}}^{\vec{k}} \partial_{\vec{\alpha}^*}^{\vec{l}} e^{\tfrac{1}{2} \vec{x}^T \Sigma \vec{x}} |_{\vec{\alpha} = \vec{0}} \nonumber \\ &= \frac{\mathcal{N}^2}{\sqrt{\mathrm{det}(\Lambda'_{uu})}} \mathrm{Haf} [ \Sigma_{\vec{k} \oplus \vec{l}} ]. 
\end{align}
Which finally gives an expression for the output probability by letting $\vec{l} = \vec{k}$
\begin{align}
\label{squeezed_transition_prob}
    \bra{\vec{k}} \mathcal{L}_T \left( \ket{\vec{r}}\bra{\vec{r}} \right) \ket{\vec{k}} = \frac{1}{\vec{j}!} \frac{\mathcal{N}^2}{\sqrt{\det\Lambda_{uu}'}} \mathrm{Haf} [ \Sigma_{\vec{k} \oplus \vec{k}} ]. 
\end{align}
The expression together with the discussion below of why only in very highly symmetric cases one can obtain PSRs for probabilities in GBS constitute the main result of this subsection.

Notice that Eq.~\eqref{squeezed_transition_prob} cannot naturally be written as a PSR. First, the RHS of the equation above depends on the inverse of square root of the determinant of $\Lambda'_{uu}$. This determinant has a polynomial dependence on the entries of $T$ but its inverse square root is not a polynomial in the entries of $T$ and thus it is not a polynomial in   $e^{i\theta}$.
Second, the matrix input of the hafnian also has a non-polynomial dependence on the entries of $T$ (cf. Eq.~\eqref{eq:kernel}).

However, if the loss can be pulled out of the unitary evolution, i.e., if 
\begin{align}\label{eq:fact}
T = U \Delta,
\end{align} where \(\Delta\) is a diagonal transmission matrix, on the one hand $E$ becomes independent of $U$ (and thus of any programmable phase $\theta$) and then so does $\det(\Lambda'_{uu})$ and moreover $\Sigma$ is only a quadratic polynomial in the entries of $T$ and $T^*$. 
When the factorization in Eq.~\eqref{eq:fact} is possible
one can notice that the elements of \(\Sigma\) %
have the same dependency in $\theta$ as in Eq.~\eqref{eq:blah}.
If we denote the number of photons in the output state vector \(\vec{j}\) as \(\sum_{i=0}^{M-1} j_i= n\), then one can expand the hafnian using its definition and obtain the following dependency
\begin{align}
    \mathrm{Haf} \left[ \Sigma_{\vec{j} \oplus \vec{j}} \right] = \sum_{m = -2n}^{2n} k_m e^{im\theta}, 
\end{align}
which means that the derivative can be obtained using a \(2n\)-order PSR.
This observation covers the case of interferometers with uniform losses $\Delta = \sqrt{\eta} \, \id$ with $\eta$ the net energy transmission of the interferometer and also lossless interferometer where $\eta=1$.

\begin{figure}[!t]
    \centering
    \includegraphics[width=0.95\linewidth]{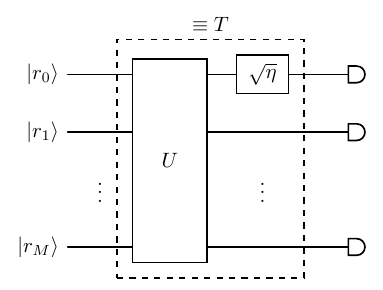}
    \caption{Example of a sub-unitary matrix with diagonalizable loss after the unitary. }
    \label{fig:example_squeezed}
\end{figure}

It may seem counter-intuitive that the PSR applies to GBS only when the loss can be commuted before the unitary evolution and not when it is applied elsewhere. To understand why, consider the $(M+1)$-mode circuit presented in Fig.~\ref{fig:example_squeezed}, where \(T=\Delta U\), and $\Delta = [\sqrt{\eta}] \oplus \id$.
Since the squeezed states have support over every even Fock state, the state after the unitary retains unbounded Fock support. Now suppose that the detector on the first output mode measures \(n\) photons. Because the loss acts after the unitary, the measured outcome receives contributions from every Fock component containing at least \(n\) photons in that mode. Consequently, the corresponding transition probability can be written as
\begin{align}
\label{GBS_impo}
    \mathrm{Pr}&[n \oplus \vec{j} | \Delta U, \vec{r}] = \nonumber \\  &\sum_{k=n}^{\infty} \binom{k}{n} \eta^n (1 - \eta)^{k-n} \cdot \mathrm{Pr}[k \oplus \vec{j}| U, \vec{r}],
\end{align}
where we used the fact that each photon in the first mode independently follows a Bernoulli process, producing a count if it passes through the loss channel with transmission~$\eta$. Therefore, the number of transmitted photons out of the possible infinite amount $k$ incident photons follows a binomial distribution. Eq.~\eqref{GBS_impo} tell us that the probability of measuring \(n\) photons in mode \(0\) depends on how many photons would have been measured in that mode had there been no loss. Since squeezed states have support over all even Fock states, every photon number \(k \geq n\) from the lossless probability distribution $\mathrm{Pr}[k \oplus \vec{j}| U, \vec{r}]$ contributes to the measured lossy probability, resulting in an infinite number of contributions. Since the sum extends over arbitrarily large photon numbers, the resulting probability contains infinitely many distinct powers of \(e^{i\theta}\). Consequently, it cannot, in general, be expressed as a finite Fourier series, which is required for the PSR to work. On the other hand, as we have shown, this issue does not arise when the loss is applied before the unitary, as in this case one can still imagine that single-mode lossy states enter into a perfect unitary where all the interference happens.

As we did previously for Fock states, one may be tempted to extend the result to threshold detectors~\cite{quesada2018gaussian}. The threshold detection probability can be written as
\begin{align}
    \mathrm{Tr}\!\left(\mathcal{L}_T(\rho)\,\Pi_{\vec{d}}\right)
    =
    \sum_{\vec{k}\in\mathcal{G}(\vec{d},n)}
    \bra{\vec{k}}
    \mathcal{L}_T(\rho)
    \ket{\vec{k}},
\end{align}
where \(n\) is the total number of input photons. This expression is well defined for Fock states because only finitely many Fock basis elements contribute to the sum. However, for squeezed states, which have unbounded Fock support, the threshold detection probability receives contributions from infinitely many Fock states. Consequently, the above decomposition no longer consists of a finite sum. 

In all the previous cases, we have shown that whenever an observable of a quantum system can be expressed as a polynomial matrix function whose input matrix has finite \(e^{i\theta}\) dependence, its derivative can be exactly reconstructed using a PSR, thus our results easily generalize to any non-displaced gaussian state~\cite{cardin2024photon}.

\section{Experiments} \label{sec:experiments}

\begin{figure}[!t]
    \centering
    \includegraphics[width=0.99\linewidth]{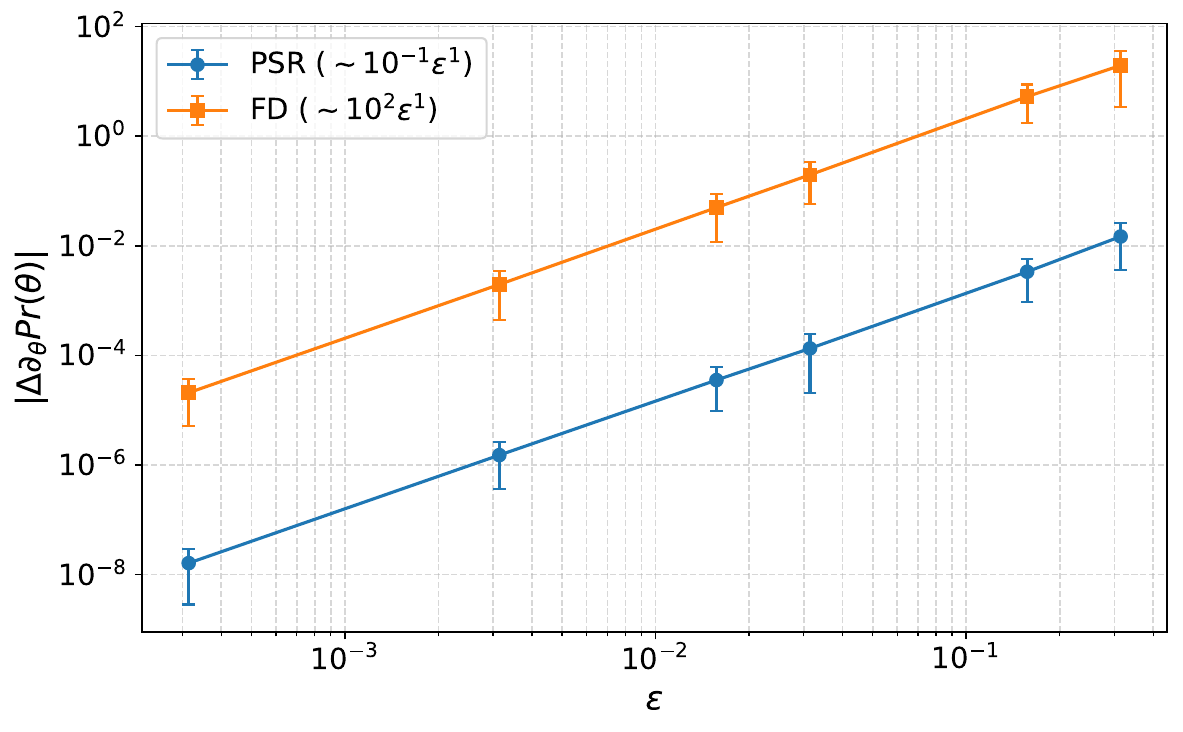}
    \caption{Absolute derivative error as a function of the phase error amplitude \(\varepsilon\) for FD with \(\delta = 10^{-4}\) (orange) and the PSR (blue). Results are shown for a four-mode random optical component with input state \(\vec{i}=(1,1,1,1)\) and output state \(\vec{j}=(2,0,2,0)\) using the SLOS model~\cite{heurtel2023strong}. Each data point represents the mean over 100 independent trials, with error bars indicating one standard deviation.}
    \label{fig:phase_error}
\end{figure}

In this section we describe the simulations and experiments performed in order to validate our approach. 

Beside loss, an  important source of error arises from imperfect phase-shifters. In realistic devices, the applied phase is subject to shot-to-shot fluctuations that can be modelled probabilistically. We describe this phase noise using a random variable where the angle implemented shot is~\cite{hamerly2022asymptotically, zhang2025fast}
\begin{align}
\theta = \bar{\theta} + \Delta \theta
\end{align}
where $\Delta \theta$ is distributed normally with mean 0 and variance $\varepsilon^2$. Phase imprecision stability is displayed in Fig.~\ref{fig:phase_error}. As one can observe, the PSR has three orders of magnitude higher tolerance to phase error than the FD method. This confirms the well-known higher tolerance to phase imprecision of the PSR over the FD method. 

\begin{figure*}[!htbp]
    \centering
    \includegraphics[width=0.99\linewidth]{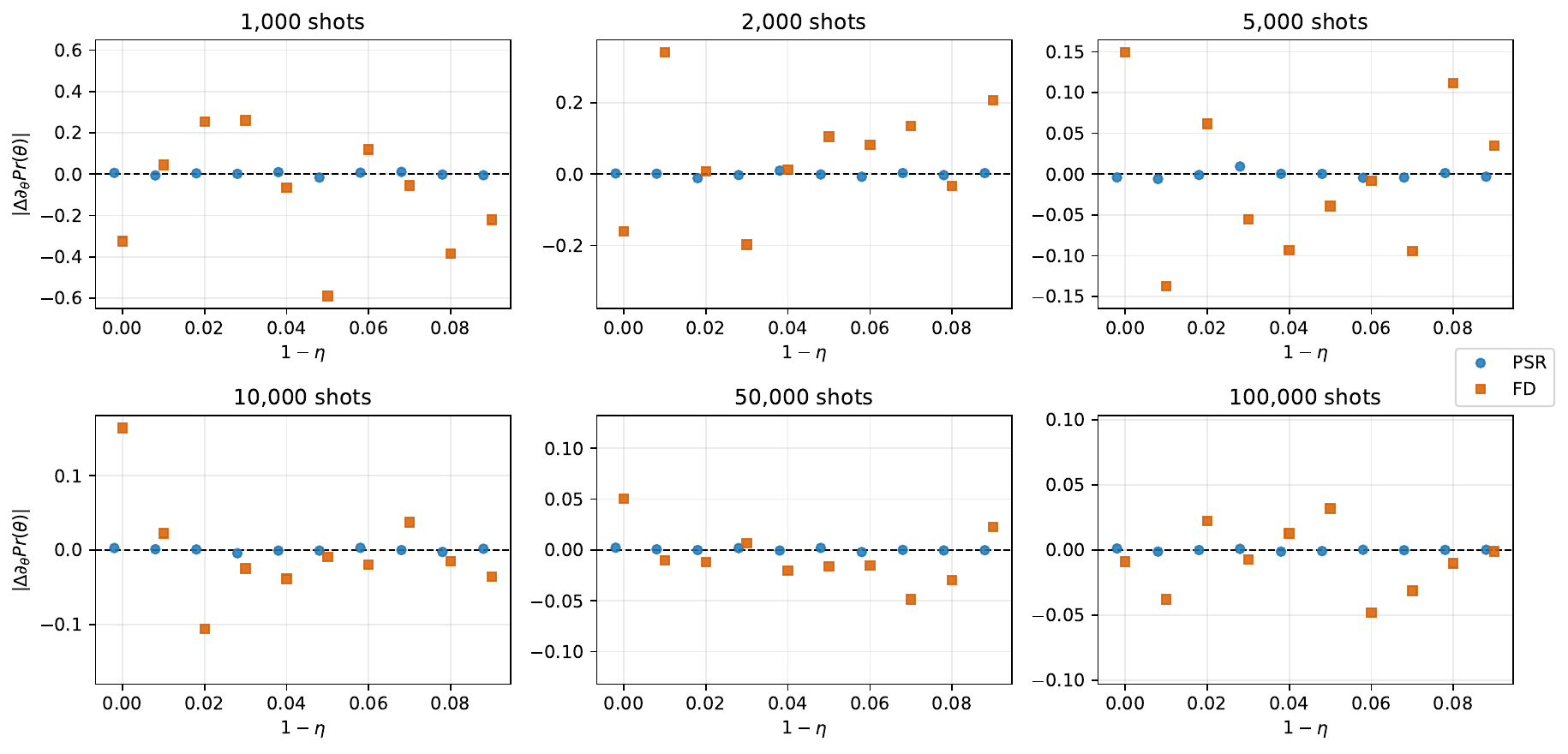}
    \caption{Derivative estimation error as a function of loss for FD with $\delta = 10^{-4}$ (orange) and the PSR (blue). Results are shown for a four-mode random optical component with input state $\vec{i}=(1,1,1,1)$ and output state $\vec{j}=(2,0,2,0)$, using Perceval's uniform noise model. Each data point corresponds to the mean derivative error over 10 independent trials.}
    \label{fig:gradient_deviations}
\end{figure*}

We can also compare the performance of the PSR with the performance of the FD method, using Perceval's~\cite{heurtel2023perceval} standard uniform loss model. Since the two methods are evaluated under the same optical losses, we expect uniform loss by itself to have a negligible effect on their relative performance. However, Perceval's noise model also includes phase errors, which have a much more pronounced impact on gradient estimation. In particular, FD is highly sensitive to small phase fluctuations because it relies on the difference of two nearly identical probability estimates, making it especially susceptible to noise. Fig.~\ref{fig:gradient_deviations} compares the deviation between the theoretical gradient and the experimentally estimated gradient for PSR and FD, as a function of the loss and for different numbers of shots (photons sent). As expected, among the two methods, the PSR consistently maintains good performance regardless of the loss level, while the FD methods exhibit larger errors and significantly higher variability. Increasing the number of shots does improve the overall performance by reducing the effect of sampling noise, but the PSR remains highly reliable even at low shot counts, highlighting its robustness.

\begin{figure}[!ht]
    \centering
    \includegraphics[width=0.99\linewidth]{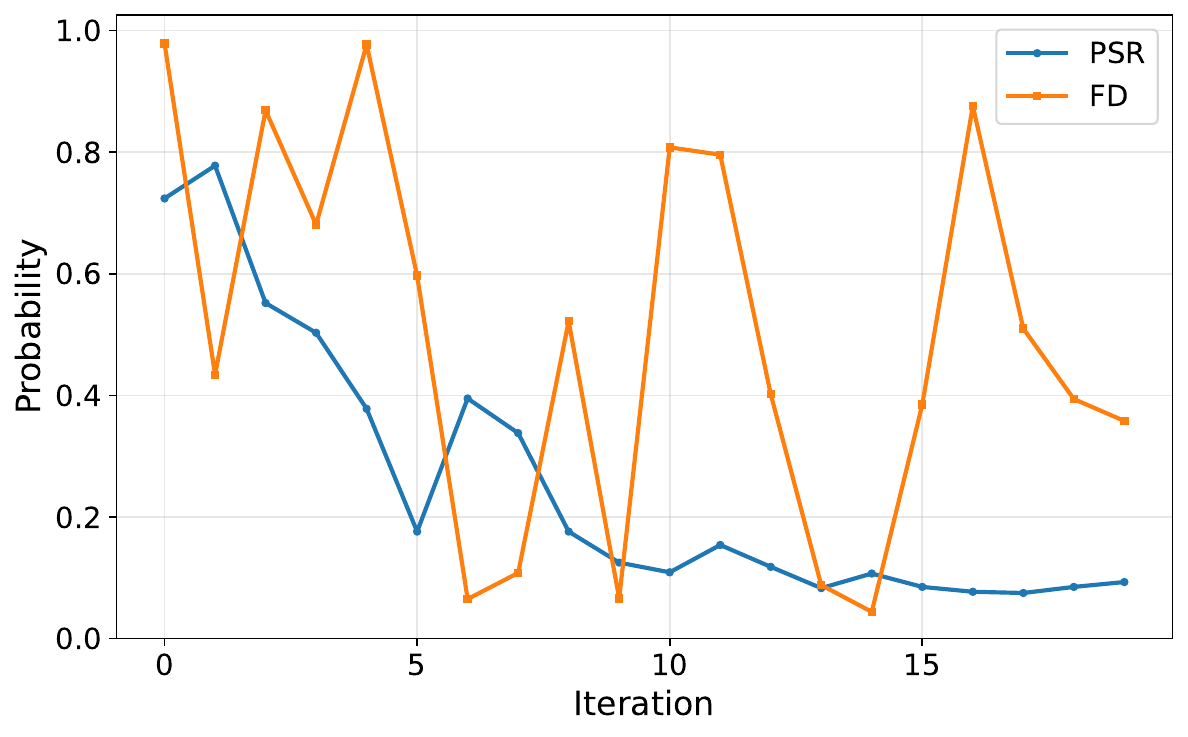}
    \caption{Cost function, corresponding to the coincidence probability $\mathrm{Pr}(\vec{j}=(1,1))$, as a function of gradient descent iterations on Quandela's Belenos QPU. The experiment was performed for two modes with initial state \(\vec{i} = (1, 1)\) and a uniformly chosen initial parameter angle and a specified shot count of 1000. The optimization was performed using a fixed step size of 0.1 for 25 iterations for FD with \(\delta = 10^{-4}\) (orange) and the PSR (blue).}
    \label{fig:grad_desc}
\end{figure}

The minimization of the coincidence probability in the Hong--Ou--Mandel~\cite{hong1987measurement} experiment was also performed on Belenos, one of Quandela's quantum processing units (QPUs). The optimization was carried out using standard gradient descent with both the PSR and FD gradient estimators. At each iteration, the gradient was estimated from photon-counting measurements obtained directly from the QPU and used to update the interferometer parameters. This provides a direct validation of the PSR in a realistic experimental setting, where sampling noise and hardware imperfections are present. Fig.~\ref{fig:grad_desc} shows the cost function optimization curves for the two methods. While both methods are able to modify the cost function during optimization, only the PSR converged toward the expected minimum coincidence probability. In contrast, the FD method exhibits extremely unstable optimization behavior due to its increased sensitivity to phase-shift noise.

\section{Discussion and Conclusion}\label{sec:end}

In this work, we derived PSRs for the transition probability from an input Fock state \(\vec{i}\) to a measured final state \(\vec{j}\), corresponding to an array of photon counts, under arbitrary loss, going further and beyond previous works~\cite{pappalardo2024photonic,facelli2024exact}. We showed that the order of the PSR is given by the number of input photons sent into the interferometer \(n = |\vec{i}|\). This method generalizes to arbitrary loss in the system and remains stable under phase-shifter imprecision. We also derived exact expression for the probability of measuring \(\vec{j}\) when single-mode squeezed states are sent into a lossy interferometer. We argued that in general there is no exact finite-order parameter-shift that can be used to estimate gradients of probabilities in this case as the Fock support of squeezed states is unbounded. For the special and restrictive case of loss that can be factored into the input of the interferometer we show an explicit PSR for GBS with order \(n = 2|\vec{j}|\).  We have also generalized our result to threshold detector output states when possible. 

The performance of the Fock-state PSR was evaluated through numerical simulations and experimentally validated on a QPU. The results demonstrate that the PSR is a reliable method for estimating gradients of photonic transition probabilities, with improved robustness compared to FD.

Future work could explore extensions of these PSRs to arbitrary Gaussian states and more general loss models. In particular, alternative approaches for computing derivatives in the presence of arbitrary loss should be investigated. It would be valuable to develop methods that remain valid for arbitrary sub-unitary evolution or that explore truncations in the Fock space of Gaussian states~\cite{upreti2026exponentially}.

\section*{Acknowledgments} M.T. thanks the Mitacs Accelerate program and the NSERC CREATE Iqucode program for their financial support. %
The authors also thank S. Poveda Hospital and V. Girouard for insightful discussions and Y. Cardin for providing feedback on the manuscript. The authors thank NSERC of Canada for financial support.

\bibliography{bib.bib}
\bibliographystyle{IEEEtran}

\end{document}